\documentclass[conference]{IEEEtran}
\IEEEoverridecommandlockouts
\usepackage{cite}
\usepackage{amsmath,amssymb,amsfonts}
\usepackage{algorithmic}
\usepackage{graphicx}
\usepackage{textcomp}
\usepackage{xcolor}
\usepackage{multirow}
\usepackage{tabularx}
\def\BibTeX{{\rm B\kern-.05em{\sc i\kern-.025em b}\kern-.08em
    T\kern-.1667em\lower.7ex\hbox{E}\kern-.125emX}}

\newcolumntype{L}[1]{>{\raggedright\arraybackslash}p{#1}}
\newcolumntype{C}[1]{>{\centering\arraybackslash}p{#1}}
\newcolumntype{R}[1]{>{\raggedleft\arraybackslash}p{#1}}

\begin{document}

% Possible conferences: 
% - https://bidma.cpsc.ucalgary.ca/IEEE-BIBM-2023/
% Possible journals:
% - International Journal of Medical Informatics (A1)

\title{Towards Automated Semantic Segmentation in Mammography Images}

\author{\IEEEauthorblockN{Cesar A. Sierra-Franco, Jan Hurtado, Victor de A. Thomaz, \\ Leonardo C. da Cruz, Santiago V. Silva, and Alberto B. Raposo}
\IEEEauthorblockA{\textit{Department of Informatics and Tecgraf Institute}\\ \textit{Pontifícia Universidade Católica do Rio de Janeiro}\\ Rio de Janeiro, Brazil\\ \{casfranco,hurtado,victorthomaz,lccruz,santiagosvs,abraposo\}@tecgraf.puc-rio.br}}

\maketitle

\begin{abstract}
Mammography images are widely used to detect non-palpable breast lesions or nodules, preventing cancer and providing the opportunity to plan interventions when necessary. The identification of some structures of interest is essential to make a diagnosis and evaluate image adequacy. Thus, computer-aided detection systems can be helpful in assisting medical interpretation by automatically segmenting these landmark structures. In this paper, we propose a deep learning-based framework for the segmentation of the nipple, the pectoral muscle, the fibroglandular tissue, and the fatty tissue on standard-view mammography images. We introduce a large private segmentation dataset and extensive experiments considering different deep-learning model architectures. Our experiments demonstrate accurate segmentation performance on variate and challenging cases, showing that this framework can be integrated into clinical practice. 
\end{abstract}


\section{Introduction}
\label{sec:introduction}

Mammography is a type of exam based on X-rays, whose main objective is to evaluate the breast tissue. This type of exam is highly recommended to prevent breast cancer because it is useful for visually identifying non-palpable lesions or nodules. Generally, mammography exams include two types of view, i.e., Medio-Lateral Oblique (MLO) and Cranio-Caudal (CC), applied to both breasts. These modalities present top-bottom and side views of the breast, allowing a multi-perspective simultaneous analysis. Both views capture a set of anatomical structures of interest that are useful to locate abnormalities and evaluate image acquisition adequacy. The reference structures consider but are not limited to the nipple, the pectoral muscle, the fibro-glandular tissue, and the fatty tissue. The automatic segmentation of these structures can assist the medical interpretation or help less experienced operators to understand the image content. 

In this paper, we introduce a study for the application of deep learning models on the mammography image segmentation task. More precisely, on the segmentation of the structures of interest mentioned above and considering MLO and CC views. We present a large private dataset that was created for this purpose and experiments with different segmentation models. Our experiments show that it is possible to automate the segmentation task on mammography images with acceptable performance. 
%We also present some challenges that we found during our development that can be addressed in future work.

The rest of this paper is structured as follows. Section \ref{sec:related_work} explains some related work for mammography image segmentation. Section \ref{sec:dataset} introduces the dataset used for the segmentation task. Section \ref{sec:semantic_segmentation} shows our experiments and results using deep learning models for segmentation. Finally, Section \ref{sec:conclusion} presents our conclusion and future work.

\section{Related work}
\label{sec:related_work}

Although the segmentation task on mammography images is typically related to the segmentation of abnormalities \cite{michael2021breast}, such as masses or nodules, we focus on those methods that estimate landmark structures useful to define the spatial distribution of the breast tissues.  

The pectoral muscle is an important landmark structure used in the MLO view assessment to define the risk of abnormalities or the adequacy of the image. This structure is usually presented in the corner of the image as a triangular shape. Its segmentation is challenging because it can present variate shapes due to the multiple anatomical conditions, the occlusion caused by the fibroglandular tissue, the inclusion of the minor pectoral muscle, and the inclusion of skin folds, among others. Several methods were proposed in the literature considering conventional signal processing and statistical analysis \cite{mustra2013robust,liu2014pectoral,oliver2014one,sreedevi2015novel,taghanaki2017geometry,vikhe2017detection,rampun2017fully,hazarika2018novel,toz2018single,ahmed2020images,divyashree2022segmentation}. Rampun et al. propose a multi-step method that uses a deep learning model for pectoral muscle boundary segmentation and post-processing steps to obtain an accurate delimitation \cite{rampun2019breast}. Similarly, Soleimani et al. propose a deep learning model to segment the pectoral muscle boundary and a graph-based analysis to improve it \cite{soleimani2020segmentation}. Instead of segmenting the boundary, Ali et al. present full pectoral muscle shape segmentation using a U-Net-based deep learning model \cite{ali2020enhancing}. Also focusing on the segmentation of the full shape of the pectoral muscle, Guo et al. propose a two-step method that uses a U-Net to estimate pectoral muscle confident regions and a GAN to estimate the final pectoral muscle shape \cite{guo2020automatic}. Rubio and Montiel present a comparative study considering multiple deep learning models and metrics for pectoral muscle segmentation \cite{rubio2021multicriteria}. In addition to the pectoral muscle, they also consider the breast shape for segmentation. Yu et al. propose a novel deep learning model that includes an attention mechanism to achieve better results than simple encoder-decoder models.  

Differently from the MLO view, the pectoral muscle in the CC view is not always present, and its recognition is more challenging. The delimitation of the pectoral muscle in the CC view is essential because it serves as a reference for the posterior limit of the breast tissues. However, its automated segmentation did not receive enough attention in the literature. Ge et al. propose a shape-guided region-growing method that uses ellipse fitting to approximate the pectoral muscle shape \cite{ge2011automatic}. This approach is redesigned and improved in \cite{ge2014segmenting} by using Markov random fields.

The nipple is another important landmark structure because it is useful to register multiple views or modalities. This allows the operator to match regions of interest and perform anatomical measurements. Some methods rely on shape and texture analysis over different regions of the breast boundary \cite{yin1994computerized,mendez1996automatic,chandrasekhar1997simple,mustra2009nipple}. The methods proposed in \cite{zhou2004computerized} and \cite{kinoshita2008radon} are based on the assumption that the fibroglandular tissue converges at the nipple position. Thus, the authors designed geometric descriptors to find the optimal convergence point. Casti et al. propose a Hessian-based method that considers geometric descriptors and constraints to define the nipple position \cite{casti2013automatic}. Jiang et al. propose a random forest classifier to detect subtle nipples using quantitative radiomic features to define regions of interest \cite{jiang2019radiomic}. Lin et al. propose a deep learning classification model applied to a set of candidate patches extracted from the mammography image  \cite{lin2019nipple}. The region that presents more intersections of patches classified as possible nipples is selected as the nipple position. These methods focus on defining a specific nipple position instead of segmenting it.

The fibroglandular tissue is a risk region that requires a special focus during the medical evaluation. Depending on the patient's anatomy, this tissue can be dense or scattered, where higher density presents a higher risk. Several methods were proposed in the literature to segment dense fibroglandular tissue regions \cite{he2015review}, considering handcrafted \cite{matsubara2001automated,el2010expectation,highnam2010robust,torres2019morphological} and data-driven \cite{keller2012estimation,keller2015preliminary,saffari2020fully,larroza2022breast,hu2022breast} models. Although dense regions are the most critical, clinical experts should also take care of scattered areas where abnormalities can be found. Thus, segmenting dense and scattered regions is important for the spatial description of the breast.

Other methods integrate the segmentation of different landmark structures in a single solution. Tiryaki et al. present some experiments using multiple U-Net-based models for the segmentation of the pectoral muscle, dense fibroglandular tissue regions, and fatty tissues \cite{tiryaki2022deep}. Considering these structures and adding the nipple, Dubrovina et al. introduce a novel deep learning-based framework for the segmentation task \cite{dubrovina2018computational}. Using multiple deep learning models, Bou presents segmentation results considering more granular structures, such as vessels, calcifications, and skin, among others \cite{bou2019deep}. All of these methods consider small datasets, making their evaluations less robust and confident for their consideration in a real-world application. Also, the first two focus on the MLO view only, as most of the segmentation methods described in this section.

This paper presents a method for the integrated segmentation of the main landmark structures, i.e., the pectoral muscle, the nipple, the fibroglandular tissue, and the fatty tissue. Unlike most segmentation methods, we focus on both standard views, MLO and CC. Thus, we are incorporating novel solutions for the segmentation of the pectoral muscle and the nipple on CC view images. In the case of fibroglandular tissue segmentation, our method considers dense and scattered areas as regions of interest, making it useful for spatial description instead of density analysis.      
Also, to the best of our knowledge, we are introducing the largest dataset for the segmentation of multiple landmark structures in mammography images. The latter allows us to show robust experiments using multiple deep-learning models. 

\section{Dataset}
\label{sec:dataset}

\begin{table*}[t]
    \centering
    \caption{Data stratification subsets}
    \begin{tabular}{|C{1.5cm}|C{0.5cm}|C{0.5cm}|C{0.5cm}|C{0.5cm}|C{0.5cm}|C{0.5cm}|C{0.5cm}|C{0.5cm}|C{0.5cm}|C{0.5cm}|C{0.5cm}|C{0.5cm}|}
    \hline
    \multicolumn{1}{|c|}{\multirow{2}{*}{\textbf{Subset}}}&\multicolumn{6}{c|}{\textbf{MLO}}&\multicolumn{6}{c|}{\textbf{CC}}\\
    \cline{2-13} 
    &\textbf{A}&\textbf{B}&\textbf{C}&\textbf{D}&\textbf{N/D}&\textbf{Total}&\textbf{A}&\textbf{B}&\textbf{C}&\textbf{D}&\textbf{N/D}&\textbf{Total}\\
    \cline{1-13}
    \textbf{Train}&903&663&347&14&1523&3450&1000&810&502&64&1361&3737\\
    \cline{1-13}
    \textbf{Validation}&209&231&229&130&407&1206&133&119&112&108&471&943\\
    \cline{1-13}
    \textbf{Test}&84&96&95&90&192&557&47&57&55&58&240&457\\
    \hline
    \end{tabular}
    \label{tab:subsets}
\end{table*}

\subsection{Raw data}

To construct a mammography image segmentation dataset, we collected a set of 2581 mammography examinations performed in a specialized hospital using different General Electric equipment. These examinations include 5213 MLO view images and 5137 CC view images in DICOM format, where each examination can repeat or lack any of the views for any of the patient's breasts. All the images were acquired using full-field digital mammography technology and present pixel spacing equal to 0.1mm or 0.094mm.

This raw data do not consider images that present breast implants, complex surgeries, or image artifacts that complicate the visualization of breast tissues. Some cases with abnormalities are included but without a corresponding clinical evaluation and categorization.

\subsection{Annotation guidelines}

We consider four landmark structures for the segmentation problem, i.e., nipple, pectoral muscle, fibroglandular tissue, and fatty tissue. Because we need a consistent segmentation dataset, we defined a set of guidelines for the delimitation of these structures that was discussed with clinical experts.

For the nipple structure, we focus on the external and internal nipple tissues, avoiding the inclusion of the areola. If the nipple location is not clear, we can consider the fibroglandular tissue convergence region to guide a deeper assessment. The external limit can be easily differentiated from the image background in cases where the nipple is in profile and is not inverted or flat. For inverted and flat in-profile nipples, we define an external limit parallel to the breast contour. In both scenarios, the internal limit is defined on the intersection with the milk ducts. The delimitation of nipples that are not in profile presents a higher complexity. In this case, we delineate the bright region generated by the nipple overlapping projection. This region is usually presented as a clear rounded shape, but it can be confused with fibroglandular tissue or abnormalities, needing clinical expert assistance in some cases. The same methodology is applied in both standard views.

The pectoral muscle in the MLO view usually presents a bright triangular shape with noticeable muscle texture and certain curvature, where the boundary with the fatty tissue is mostly clear. However, in cases where the fibroglandular tissue is dense, the bottom region of the pectoral muscle can be occluded, making its delineation difficult. Thus, we delineate the clear boundary and its corresponding projection on the diffuse regions.

In the CC view, the pectoral muscle is not always present and can be in different sizes. We delineate this structure if the shape is visible by manipulating the image contrast and if it represents a considerable area within the image. We avoid the inclusion of very thin structures that can be confused with an image boundary artifact. This is the most challenging structure to segment, requiring intensive clinical expert support.

Our fibroglandular tissue delineation is not limited to the segmentation of dense regions; we also include scattered and fatty regions where we can find agglomeration of ducts or glands. Thus, our structure represents a region of interest for the fibroglandular tissue instead of an accurate selection of tissues. 

We consider fatty tissue every breast content that is outside the structures defined previously. Its segmentation can be simplified by just delineating the breast boundary, including a portion of the abdomen, and then subtracting the other structure shapes automatically.  

\subsection{Annotation tool}

We implemented a sketch-based contour drawing annotation tool to delineate the structures of interest. This tool allows the user to draw a closed contour for a given structure and edit it if necessary, considering an intuitive interaction for deformation. The user can also manipulate the image by adjusting the standard window/level parameters to obtain a clearer visualization and by zooming and translating it to focus on the target structure. The contours are saved as dense high-resolution polygons defined in the image space.
%A conventional polygon editing tool can cause several delay in the annotation process because the annotations require several polygon vertices.

The annotation tool allows an automatic initialization of the breast contour for the fatty tissue to speed up the annotation process. This breast contour initialization is computed by segmenting the image using the Otsu binary thresholding \cite{otsu1979threshold} and then computing the external contour of the largest connected component.

For the other structures, the tool includes deep learning-based initializations performed by two segmentation models trained on partial versions of the annotated dataset. These models follow the same idea presented in Section \ref{sec:semantic_segmentation}, considering each standard view as an independent domain. Because these initializations are predictions, they should be edited to match the appropriate boundaries.

\subsection{Annotation process}

A team of 8 annotators was trained by two clinical experts for the identification and delimitation of the structures of interest. These clinical experts resolved the annotators' doubts about challenging cases during the entire annotation process. Unlike identifying abnormalities that require high expertise, the structures of interest for our problem are relatively easy to identify in most cases.

The raw data was split and assigned to the annotators, such that each image was annotated by a single person. The entire annotation process was realized in approximately one year, including multiple refinements based on the feedback of the clinical experts. 

At the end of the annotation process, the four structures of interest were annotated over all the 5213 MLO images. The nipple, the fibroglandular tissue, and the fatty tissue were annotated over all the 5137 CC view images. Because the pectoral muscle is not always present in this view, this structure was annotated on 2952 (57\%) of these images.   

\subsection{Data stratification}

Following the standard data splitting process for machine learning methods and considering independent sets for each view, we randomly divide the annotated images into three disjoint subsets: train, validation, and test. This splitting process avoids the inclusion of samples of the same examination in different subsets and tries to balance the samples regarding the fibroglandular tissue density. The density-based balancing is possible because some of the examinations were previously categorized by clinical experts into the four standard density classes: (A) almost entirely fatty, (B) scattered fibroglandular tissue, (C) heterogeneously dense, and (D) extremely dense. However, the majority of examinations do not present this density annotation. 

The distribution of the subsets for each view is presented in Table \ref{tab:subsets}, considering the density annotations. Notice how the test subset presents a balanced behavior regarding the density classes, making it a good benchmark for evaluation.

\section{Mammography deep learning semantic segmentation}
\label{sec:semantic_segmentation}

\subsection{Overview}

\begin{figure}[t]
\centering
\includegraphics[width=1\columnwidth]{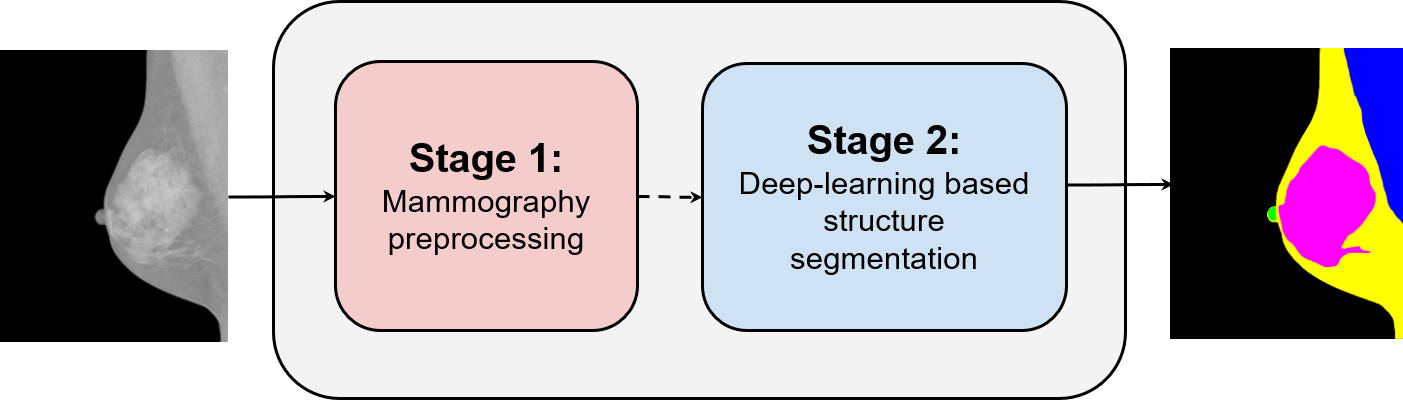}
\caption{Mammography deep learning segmentation}
\label{Fig:segmentation-pipeline}
\end{figure}

Figure \ref{Fig:segmentation-pipeline} shows the mammography image segmentation pipeline proposed in this paper. This pipeline focuses on the segmentation of the structures of interest and the image background. The proposed solution employs a deep learning-based approach using semantic segmentation neural networks. The semantic segmentation process involves predicting labels for each pixel in the image, assigning them to specific classes. This process establishes meaningful regions within the image, associating each pixel with semantically relevant information. In our solution, we consider five classes of region: background, nipple, pectoral muscle, fibroglandular tissue, and fatty tissue. The deep learning model consumes a pre-processed mammography image and generates a set of probability maps that define the degree of correspondence of each pixel to a given class.

\subsection{Evaluation metric}

To assess the segmentation accuracy of the target structures in the resulting trained models, we use the Intersection over Union (IoU) score, a widely-used metric in segmentation tasks. The IoU measures the overlap between the model-predicted structures and the ground-truth annotations. This score is calculated by dividing the intersection area between the predicted and ground-truth regions by the area of their union. This ratio yields a value between 0 and 1, where a score of 1 indicates a perfect overlap between the predicted and ground-truth structures, while a score of 0 suggests no overlap at all. Higher IoU scores indicate that the model's predictions align more closely with the desired target structures.

\subsection{Pre-processing}

To create ground-truth probability maps for each structure, we first rasterize the annotated polygons into a single multi-class label map considering the following operations: (1) fill the full label map with the background class label. (2) rasterize the fatty tissue polygon on the previous label map, i.e., the breast contour polygon. (3) rasterize the fibroglandular tissue polygon on the previous label map. (4) rasterize the pectoral muscle polygon on the previous label map. (5) rasterize the nipple polygon on the previous label map. Then, this label map is converted to the probability maps by using one-hot encoding.

The mammography image is represented in the DICOM file as an integer grayscale image with values in the range $[0,4095]$. To prepare this image for the deep-learning model consumption, we apply the following operations: (1) image normalization using the percentiles 2 and 98 as minimum and maximum values. (2) adaptive histogram equalization using the algorithm Contrast Limited Adaptive Histogram Equalization (CLAHE) \cite{clahe} with kernel size $1/8$ times the height and width of the image. (3) image intensity rescaling to the range $[0,255]$. 

To reduce the domain space, we flip all the left breast images and their corresponding probability maps horizontally. Thus, all the images are treated as right breast images.

%\subsection{Neural networks training details}
\subsection{Neural Network architectures results}

Due to the distinct characteristics of MLO and CC views, we trained separate models for each one. In our training experiments, we test several architectures for semantic segmentation, including Unet, FPN, Linknet, and PSPNet. While all these architectures serve the same purpose they employ different strategies.

%\begin{itemize}

 %   item UNet: encoder-decoder type architecture that use skip connections to facilitate the information flow between encoder and decoder layers. 

%    \item Pyramid Scene Parsing Network (PSPnet): uses a pyramid analysis module to consider the global context of the image and improve the local level forecasts.
    
 %   \item LinkNet: Network similar to Unet using an encoder-decoder type architecture with few modifications aiming to be faster, designed for real-time applications.
    
%    \item FPN:Feature Pyramid Network (FPN):  has a pyramidal structure. It performs predictions for various scales of the image, concatenating the results to obtain the final segmentation masks. 

%\end{itemize}

%The initial training hyper-parameters used for the performed experiments are listed below.

Based on extensive hyper-parameter optimization experiments, we selected the following training configuration. The model input is a single-channel $384\times384$ image with intensity values in the range $[0,1]$. Thus, the pre-processed image should be rescaled and resized to the corresponding range and size. The model output is a tensor that represents the probability maps of the structures of interest, including the background. We use softmax as the activation function in the final layer to help the model to predict a single class for each pixel, i.e., a multi-class problem. Further, because the segmentation models allow the integration of a custom convolutional backbone, we consider the EfficientnetB3 as the feature extractor.

During training, we use the Jaccard loss function with learning rate $10^{-3}$ and batch size $4$. We consider an early stopping scheme with 65 as the maximum number of epochs and 20 as patience, selecting the best weights on the validation dataset regarding the loss function.  

Tables \ref{tab:cc_mean_iou} and \ref{tab:mlo_mean_iou}  present the mean IoU results per structure for the CC and MLO view, respectively. From the tables, it is possible to observe good IoU scores above 0.7 for all the segmented structures obtained from the test set. In this mammography segmentation context, the PSPNet network presented the worst metrics, while the FPN showed the best results, improving pectoral muscle detection.

%\textcolor{red}{Completar as tabelas e a descrição -- Santiago}

\begin{table}[t]
\centering
\begin{center}
\caption{CC Mean IoU results on test set per segmented structure}
\begin{tabular}{|l|c|c|c|c|c|}
\hline
\multicolumn{1}{|c|}{\textbf{Architecture}} & \textbf{Nipple} & \textbf{\begin{tabular}[c]{@{}c@{}}Pectoral \\ muscle\end{tabular}} & \textbf{\begin{tabular}[c]{@{}c@{}}Fibro \\ tissue\end{tabular}} & \textbf{\begin{tabular}[c]{@{}c@{}}Fatty \\ tissue\end{tabular}} & \textbf{\begin{tabular}[c]{@{}c@{}}Mean \end{tabular}} \\ \hline
\textbf{Unet} & 0,77 & 0.78 & 0.88 & 0.76 & 0.84 \\ \hline
\textbf{PSPNet} & 0.75 & 0.73 & 0.88 & 0.76 & 0.83 \\ \hline
\textbf{Linknet} & 0.76 & 0.75 & 0.90 & 0.81 & 0.84 \\ \hline
\textbf{FPN} & 0.77 & 0.79 & 0.90 & 0.81 & 0.85 \\ \hline
\end{tabular}
\label{tab:cc_mean_iou}
\end{center}
\end{table}

\begin{table}[t]
\centering
\begin{center}
\caption{MLO Mean IoU results on test set per segmented structure}
\begin{tabular}{|l|c|c|c|c|c|}
\hline
\multicolumn{1}{|c|}{\textbf{Architecture}} & \textbf{Nipple} & \textbf{\begin{tabular}[c]{@{}c@{}}Pectoral \\ muscle\end{tabular}} & \textbf{\begin{tabular}[c]{@{}c@{}}Fibro \\ tissue\end{tabular}} & \textbf{\begin{tabular}[c]{@{}c@{}}Fatty \\ tissue\end{tabular}} & \textbf{\begin{tabular}[c]{@{}c@{}}Mean \end{tabular}} \\ \hline
\textbf{Unet} & 0.70 & 0.95 & 0.90 & 0.80 & 0.87 \\ \hline
\textbf{PSPNet} & 0.71 & 0.96 & 0.89 & 0.80 & 0.87 \\ \hline
\textbf{Linknet} & 0.72 & 0.95 & 0.89 & 0.79 & 0.87 \\ \hline
\textbf{FPN} & 0.73 & 0.96 & 0.91 & 0.82 & 0.88 \\ \hline
\end{tabular}
\label{tab:mlo_mean_iou}
\end{center}
\end{table}

\subsection{Visual results}

\begin{figure}[t]
\centering
\includegraphics[width=0.9\columnwidth]{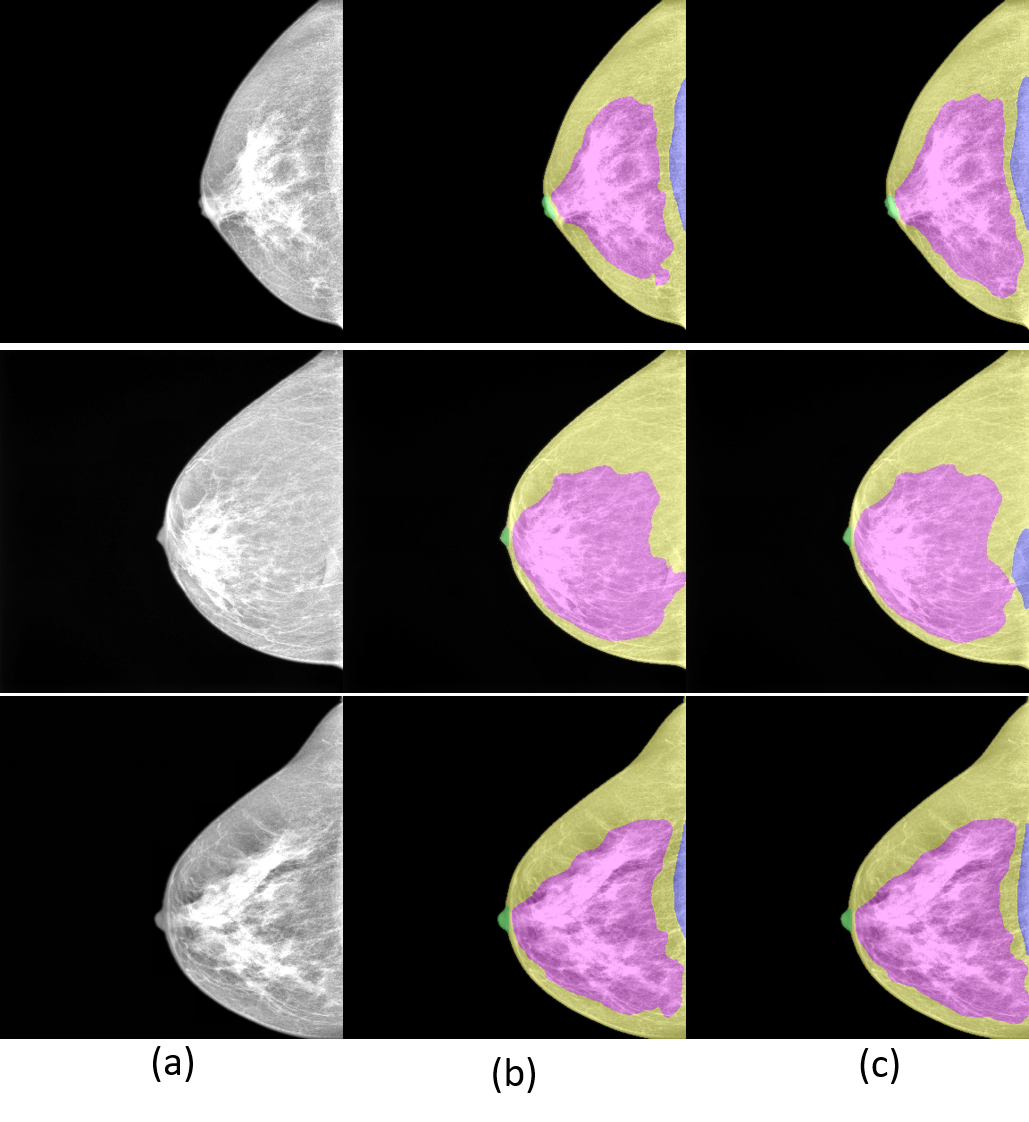}
\caption{CC model visual results: a) preprocessed image, b) ground-truth and c) model prediction }
\label{Fig:ccvisual}
\end{figure}

\begin{figure}[t]
\centering
\includegraphics[width=0.9\columnwidth]{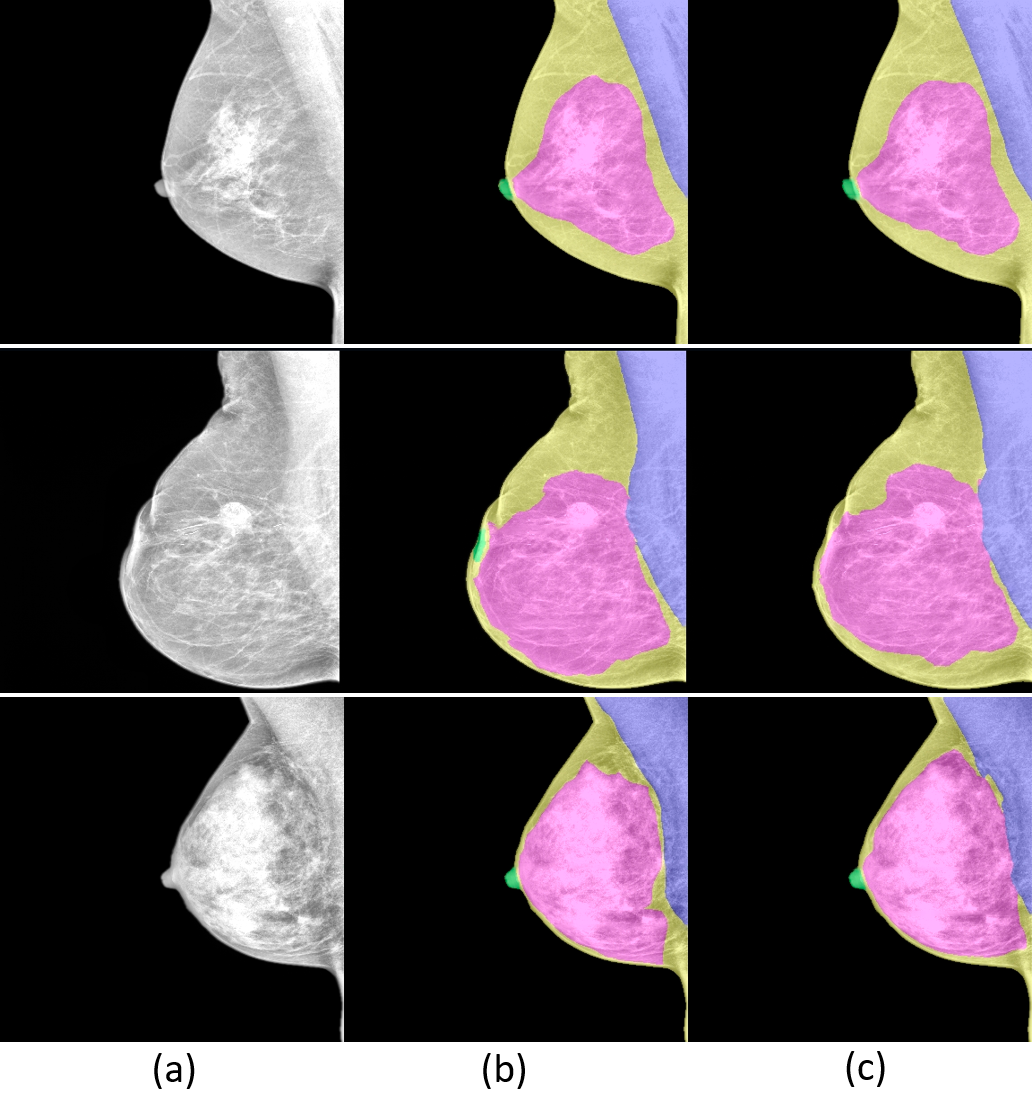}
\caption{MLO model visual results: a) preprocessed image, b) ground-truth and c) model prediction }
\label{Fig:mlovisual}
\end{figure}

Figures \ref{Fig:ccvisual} and \ref{Fig:mlovisual} illustrate some visual results for the models' segmentation predictions. In both Figures, column a) refers to the preprocessed image, column b) represents the ground-truth, and column c) the model prediction. From the visual results, it's possible to observe the effectiveness of the model's segmentation predictions. Comparing the ground truth with the model's predictions in column c), we can assess the model's performance in accurately delineating the structures of interest. These visual results offer insights into the models' ability to capture important features and accurately segment the desired mammogram structures.

Despite the positive results, upon visual inspection, we observe that there is still room for improvement. For instance, in the second row of Figure \ref{Fig:ccvisual}, there is a false positive segmentation of the pectoral muscle on the CC view. From the visual point of view, the pectoral structure is not always present in this view, and when it is present, it could not be easy to spot even for human specialists. This structure presents a high variability of cases that impact this structure's visualization. Sometimes it overlaps with the fibroglandular tissue and does not always present a clear pattern like in the MLO view. Another imprecision could be observed on some mammograms with flat or not-in-profile nipples, such as the sample illustrated on the second row in Figure \ref{Fig:mlovisual}. 
 
These observations highlight a potential area for refinement in the model's performance. While the overall results may be promising, identifying and addressing such instances of inaccuracies can contribute to enhancing the model's segmentation accuracy and overall effectiveness.

\section{Conclusion and future work}
\label{sec:conclusion}

We presented a deep learning-based solution for the segmentation of structures of interest in mammography images. This solution includes a fully-annotated dataset useful for training the models in a supervised manner. The segmentation results show that this method is at least promising for its integration into clinical practice.

Although the results are encouraging, there is still a gap for improvement in the estimation of more accurate segments, especially for the challenging structures. Further, this idea can be extended for detecting and segmenting other structures, such as skin folds, vessels, abnormalities, dense fibroglandular tissue regions, and image artifacts, among others.

%\section*{Acknowledgment}




\bibliographystyle{IEEEtran}
\bibliography{main.bib}

\begin{thebibliography}{10}
\providecommand{\url}[1]{#1}
\csname url@samestyle\endcsname
\providecommand{\newblock}{\relax}
\providecommand{\bibinfo}[2]{#2}
\providecommand{\BIBentrySTDinterwordspacing}{\spaceskip=0pt\relax}
\providecommand{\BIBentryALTinterwordstretchfactor}{4}
\providecommand{\BIBentryALTinterwordspacing}{\spaceskip=\fontdimen2\font plus
\BIBentryALTinterwordstretchfactor\fontdimen3\font minus
  \fontdimen4\font\relax}
\providecommand{\BIBforeignlanguage}[2]{{%
\expandafter\ifx\csname l@#1\endcsname\relax
\typeout{** WARNING: IEEEtran.bst: No hyphenation pattern has been}%
\typeout{** loaded for the language `#1'. Using the pattern for}%
\typeout{** the default language instead.}%
\else
\language=\csname l@#1\endcsname
\fi
#2}}
\providecommand{\BIBdecl}{\relax}
\BIBdecl

\bibitem{michael2021breast}
E.~Michael, H.~Ma, H.~Li, F.~Kulwa, and J.~Li, ``Breast cancer segmentation
  methods: current status and future potentials,'' \emph{BioMed Research
  International}, vol. 2021, pp. 1--29, 2021.

\bibitem{mustra2013robust}
M.~Mustra and M.~Grgic, ``Robust automatic breast and pectoral muscle
  segmentation from scanned mammograms,'' \emph{Signal processing}, vol.~93,
  no.~10, pp. 2817--2827, 2013.

\bibitem{liu2014pectoral}
L.~Liu, Q.~Liu, and W.~Lu, ``Pectoral muscle detection in mammograms using
  local statistical features,'' \emph{Journal of digital imaging}, vol.~27, pp.
  633--641, 2014.

\bibitem{oliver2014one}
A.~Oliver, X.~Llad{\'o}, A.~Torrent, and J.~Mart{\'\i}, ``One-shot segmentation
  of breast, pectoral muscle, and background in digitised mammograms,'' in
  \emph{2014 IEEE International Conference on Image Processing (ICIP)}.\hskip
  1em plus 0.5em minus 0.4em\relax IEEE, 2014, pp. 912--916.

\bibitem{sreedevi2015novel}
S.~Sreedevi and E.~Sherly, ``A novel approach for removal of pectoral muscles
  in digital mammogram,'' \emph{Procedia Computer Science}, vol.~46, pp.
  1724--1731, 2015.

\bibitem{taghanaki2017geometry}
S.~A. Taghanaki, Y.~Liu, B.~Miles, and G.~Hamarneh, ``Geometry-based pectoral
  muscle segmentation from mlo mammogram views,'' \emph{IEEE Transactions on
  Biomedical Engineering}, vol.~64, no.~11, pp. 2662--2671, 2017.

\bibitem{vikhe2017detection}
P.~Vikhe and V.~Thool, ``Detection and segmentation of pectoral muscle on
  mlo-view mammogram using enhancement filter,'' \emph{Journal of medical
  systems}, vol.~41, pp. 1--13, 2017.

\bibitem{rampun2017fully}
A.~Rampun, P.~J. Morrow, B.~W. Scotney, and J.~Winder, ``Fully automated breast
  boundary and pectoral muscle segmentation in mammograms,'' \emph{Artificial
  intelligence in medicine}, vol.~79, pp. 28--41, 2017.

\bibitem{hazarika2018novel}
M.~Hazarika and L.~B. Mahanta, ``A novel region growing based method to remove
  pectoral muscle from mlo mammogram images,'' in \emph{Advances in
  Electronics, Communication and Computing: ETAEERE-2016}.\hskip 1em plus 0.5em
  minus 0.4em\relax Springer, 2018, pp. 307--316.

\bibitem{toz2018single}
G.~Toz and P.~Erdogmus, ``A single sided edge marking method for detecting
  pectoral muscle in digital mammograms,'' \emph{Engineering, Technology and
  Applied Science Research}, vol.~8, no.~1, pp. 2367--2373, 2018.

\bibitem{ahmed2020images}
L.~Ahmed, M.~M. Iqbal, H.~Aldabbas, S.~Khalid, Y.~Saleem, and S.~Saeed,
  ``Images data practices for semantic segmentation of breast cancer using deep
  neural network,'' \emph{Journal of Ambient Intelligence and Humanized
  Computing}, pp. 1--17, 2020.

\bibitem{divyashree2022segmentation}
B.~Divyashree, R.~Amarnath, M.~Naveen, and H.~Kumar, ``Segmentation of pectoral
  muscle in mammograms using granular computing,'' \emph{Journal of Information
  Technology Research (JITR)}, vol.~15, no.~1, pp. 1--14, 2022.

\bibitem{rampun2019breast}
A.~Rampun, K.~L{\'o}pez-Linares, P.~J. Morrow, B.~W. Scotney, H.~Wang, I.~G.
  Oca{\~n}a, G.~Maclair, R.~Zwiggelaar, M.~A.~G. Ballester, and I.~Mac{\'\i}a,
  ``Breast pectoral muscle segmentation in mammograms using a modified
  holistically-nested edge detection network,'' \emph{Medical image analysis},
  vol.~57, pp. 1--17, 2019.

\bibitem{soleimani2020segmentation}
H.~Soleimani and O.~V. Michailovich, ``On segmentation of pectoral muscle in
  digital mammograms by means of deep learning,'' \emph{IEEE Access}, vol.~8,
  pp. 204\,173--204\,182, 2020.

\bibitem{ali2020enhancing}
M.~J. Ali, B.~Raza, A.~R. Shahid, F.~Mahmood, M.~A. Yousuf, A.~H. Dar, and
  U.~Iqbal, ``Enhancing breast pectoral muscle segmentation performance by
  using skip connections in fully convolutional network,'' \emph{International
  Journal of Imaging Systems and Technology}, vol.~30, no.~4, pp. 1108--1118,
  2020.

\bibitem{guo2020automatic}
Y.~Guo, W.~Zhao, S.~Li, Y.~Zhang, and Y.~Lu, ``Automatic segmentation of the
  pectoral muscle based on boundary identification and shape prediction,''
  \emph{Physics in Medicine \& Biology}, vol.~65, no.~4, p. 045016, 2020.

\bibitem{rubio2021multicriteria}
Y.~Rubio and O.~Montiel, ``Multicriteria evaluation of deep neural networks for
  semantic segmentation of mammographies,'' \emph{Axioms}, vol.~10, no.~3, p.
  180, 2021.

\bibitem{ge2011automatic}
M.~Ge, G.~Mawdsley, and M.~Yaffe, ``Automatic identification of pectoralis
  muscle on digital cranio-caudal-view mammograms,'' in \emph{Medical Imaging
  2011: Computer-Aided Diagnosis}, vol. 7963.\hskip 1em plus 0.5em minus
  0.4em\relax SPIE, 2011, pp. 572--579.

\bibitem{ge2014segmenting}
M.~Ge, J.~G. Mainprize, G.~E. Mawdsley, and M.~J. Yaffe, ``Segmenting
  pectoralis muscle on digital mammograms by a markov random field-maximum a
  posteriori model,'' \emph{Journal of Medical Imaging}, vol.~1, no.~3, pp.
  034\,503--034\,503, 2014.

\bibitem{yin1994computerized}
F.-F. Yin, M.~L. Giger, K.~Doi, C.~J. Vyborny, and R.~A. Schmidt,
  ``Computerized detection of masses in digital mammograms: Automated alignment
  of breast images and its effect on bilateral-subtraction technique,''
  \emph{Medical Physics}, vol.~21, no.~3, pp. 445--452, 1994.

\bibitem{mendez1996automatic}
A.~J. M{\'e}ndez, P.~G. Tahoces, M.~J. Lado, M.~Souto, J.~Correa, and J.~J.
  Vidal, ``Automatic detection of breast border and nipple in digital
  mammograms,'' \emph{Computer methods and programs in biomedicine}, vol.~49,
  no.~3, pp. 253--262, 1996.

\bibitem{chandrasekhar1997simple}
R.~Chandrasekhar and Y.~Attikiouzel, ``A simple method for automatically
  locating the nipple on mammograms,'' \emph{IEEE transactions on medical
  imaging}, vol.~16, no.~5, pp. 483--494, 1997.

\bibitem{mustra2009nipple}
M.~Mustra, J.~Bozek, and M.~Grgic, ``Nipple detection in craniocaudal digital
  mammograms,'' in \emph{2009 International Symposium ELMAR}.\hskip 1em plus
  0.5em minus 0.4em\relax IEEE, 2009, pp. 15--18.

\bibitem{zhou2004computerized}
C.~Zhou, H.-P. Chan, C.~Paramagul, M.~A. Roubidoux, B.~Sahiner, L.~M.
  Hadjiiski, and N.~Petrick, ``Computerized nipple identification for multiple
  image analysis in computer-aided diagnosis: Computerized nipple
  identification on mammograms,'' \emph{Medical Physics}, vol.~31, no.~10, pp.
  2871--2882, 2004.

\bibitem{kinoshita2008radon}
S.~K. Kinoshita, P.~M. Azevedo-Marques, R.~R. Pereira, J.~A.~H. Rodrigues, and
  R.~M. Rangayyan, ``Radon-domain detection of the nipple and the pectoral
  muscle in mammograms,'' \emph{Journal of digital imaging}, vol.~21, pp.
  37--49, 2008.

\bibitem{casti2013automatic}
P.~Casti, A.~Mencattini, M.~Salmeri, A.~Ancona, F.~F. Mangieri, M.~L. Pepe, and
  R.~M. Rangayyan, ``Automatic detection of the nipple in screen-film and
  full-field digital mammograms using a novel hessian-based method,''
  \emph{Journal of digital imaging}, vol.~26, pp. 948--957, 2013.

\bibitem{jiang2019radiomic}
J.~Jiang, Y.~Zhang, Y.~Lu, Y.~Guo, and H.~Chen, ``A radiomic feature--based
  nipple detection algorithm on digital mammography,'' \emph{Medical physics},
  vol.~46, no.~10, pp. 4381--4391, 2019.

\bibitem{lin2019nipple}
Y.~Lin, M.~Li, S.~Chen, L.~Yu, and F.~Ma, ``Nipple detection in mammogram using
  a new convolutional neural network architecture,'' in \emph{2019 12th
  International Congress on Image and Signal Processing, BioMedical Engineering
  and Informatics (CISP-BMEI)}.\hskip 1em plus 0.5em minus 0.4em\relax IEEE,
  2019, pp. 1--6.

\bibitem{he2015review}
W.~He, A.~Juette, E.~R. Denton, A.~Oliver, R.~Mart{\'\i}, R.~Zwiggelaar
  \emph{et~al.}, ``A review on automatic mammographic density and parenchymal
  segmentation,'' \emph{International journal of breast cancer}, vol. 2015,
  2015.

\bibitem{matsubara2001automated}
T.~Matsubara, D.~Yamazaki, M.~Kato, T.~Hara, H.~Fujita, T.~Iwase, and T.~Endo,
  ``An automated classification scheme for mammograms based on amount and
  distribution of fibroglandular breast tissue density,'' in
  \emph{International congress series}, vol. 1230.\hskip 1em plus 0.5em minus
  0.4em\relax Elsevier, 2001, pp. 545--552.

\bibitem{el2010expectation}
A.~El-Zaart, ``Expectation--maximization technique for fibro-glandular discs
  detection in mammography images,'' \emph{Computers in Biology and Medicine},
  vol.~40, no.~4, pp. 392--401, 2010.

\bibitem{highnam2010robust}
R.~Highnam, S.~M. Brady, M.~J. Yaffe, N.~Karssemeijer, and J.~Harvey, ``Robust
  breast composition measurement-volpara tm,'' in \emph{Digital Mammography:
  10th International Workshop, IWDM 2010, Girona, Catalonia, Spain, June 16-18,
  2010. Proceedings 10}.\hskip 1em plus 0.5em minus 0.4em\relax Springer, 2010,
  pp. 342--349.

\bibitem{torres2019morphological}
G.~F. Torres, A.~Sassi, O.~Arponen, K.~Holli-Helenius, A.-L. L{\"a}{\"a}peri,
  I.~Rinta-Kiikka, J.~K{\"a}m{\"a}r{\"a}inen, and S.~Pertuz, ``Morphological
  area gradient: System-independent dense tissue segmentation in mammography
  images,'' in \emph{2019 41st Annual International Conference of the IEEE
  Engineering in Medicine and Biology Society (EMBC)}.\hskip 1em plus 0.5em
  minus 0.4em\relax IEEE, 2019, pp. 4855--4858.

\bibitem{keller2012estimation}
B.~M. Keller, D.~L. Nathan, Y.~Wang, Y.~Zheng, J.~C. Gee, E.~F. Conant, and
  D.~Kontos, ``Estimation of breast percent density in raw and processed full
  field digital mammography images via adaptive fuzzy c-means clustering and
  support vector machine segmentation,'' \emph{Medical physics}, vol.~39,
  no.~8, pp. 4903--4917, 2012.

\bibitem{keller2015preliminary}
B.~M. Keller, J.~Chen, D.~Daye, E.~F. Conant, and D.~Kontos, ``Preliminary
  evaluation of the publicly available laboratory for breast radiodensity
  assessment (libra) software tool: comparison of fully automated area and
  volumetric density measures in a case--control study with digital
  mammography,'' \emph{Breast cancer research}, vol.~17, pp. 1--17, 2015.

\bibitem{saffari2020fully}
N.~Saffari, H.~A. Rashwan, M.~Abdel-Nasser, V.~Kumar~Singh, M.~Arenas,
  E.~Mangina, B.~Herrera, and D.~Puig, ``Fully automated breast density
  segmentation and classification using deep learning,'' \emph{Diagnostics},
  vol.~10, no.~11, p. 988, 2020.

\bibitem{larroza2022breast}
A.~Larroza, F.~J. P{\'e}rez-Benito, J.-C. Perez-Cortes, M.~Rom{\'a}n,
  M.~Poll{\'a}n, B.~P{\'e}rez-G{\'o}mez, D.~Salas-Trejo, M.~Casals, and
  R.~Llobet, ``Breast dense tissue segmentation with noisy labels: A hybrid
  threshold-based and mask-based approach,'' \emph{Diagnostics}, vol.~12,
  no.~8, p. 1822, 2022.

\bibitem{hu2022breast}
J.~Hu, Z.~Liu, and Q.~Wang, ``Breast density segmentation in mammograms based
  on dual attention mechanism,'' in \emph{Proceedings of the 3rd International
  Symposium on Artificial Intelligence for Medicine Sciences}, 2022, pp.
  430--435.

\bibitem{tiryaki2022deep}
V.~Tiryaki and V.~Kaplano{\u{g}}lu, ``Deep learning-based multi-label tissue
  segmentation and density assessment from mammograms,'' \emph{IRBM}, vol.~43,
  no.~6, pp. 538--548, 2022.

\bibitem{dubrovina2018computational}
A.~Dubrovina, P.~Kisilev, B.~Ginsburg, S.~Hashoul, and R.~Kimmel,
  ``Computational mammography using deep neural networks,'' \emph{Computer
  Methods in Biomechanics and Biomedical Engineering: Imaging \&
  Visualization}, vol.~6, no.~3, pp. 243--247, 2018.

\bibitem{bou2019deep}
A.~Bou, ``Deep learning models for semantic segmentation of mammography
  screenings,'' 2019.

\bibitem{otsu1979threshold}
N.~Otsu, ``A threshold selection method from gray-level histograms,''
  \emph{IEEE transactions on systems, man, and cybernetics}, vol.~9, no.~1, pp.
  62--66, 1979.

\bibitem{clahe}
K.~Zuiderveld, \emph{Contrast Limited Adaptive Histogram Equalization}.\hskip
  1em plus 0.5em minus 0.4em\relax USA: Academic Press Professional, Inc.,
  1994, p. 474–485.

\end{thebibliography}

\end{document}